\documentclass[pra]{revtex4}
\usepackage{amsmath}
\usepackage{graphicx}
\begin{document}

\title{Upper and lower bounds on the mean square radius and criteria for occurrence of quantum halo states}
\author{Fabian Brau}\email{fabian.brau@umh.ac.be}
\affiliation{Groupe de Physique Nucl\'eaire Th\'eorique, Acad\'emie Universitaire Wallonie-Bruxelles, Universit\'e de Mons-Hainaut, B-7000 Mons, Belgique}

\begin{abstract}
In the context of non-relativistic quantum mechanics, we obtain several upper and lower limits on the mean square radius applicable to systems composed by two-body bound by a central potential. A lower limit on the mean square radius is used to obtain a simple criteria for the occurrence of S-wave quantum halo sates.
\end{abstract}

\date{\today}

\maketitle

\section{Introduction}
\label{sec1}

The root mean square (rms) radius is used in many fields of physics to characterize the size of quantum systems; this observable is thus of special interest. In the case of two-body systems, the rms radius and the energy of one eigenstate determine the depth and the range of the central potential which binds the particles provided that the number of bound states supported by the potential is also known. A simple example is the naive description of the deuteron by a square-well potential (see for example \cite{brow76}). The depth $V_0$ and the range $R$ can be adjusted to reproduce the rms radius and the binding energy of the deuteron. This adjustment is not unique excepted if the number of bound states in the potential is fixed to one. Obviously, these quantities are not sufficient to infer the shape of the potential. Indeed, the same simple description of the deuteron could be achieved with an exponential potential for example. However, if information on the shape of the potential is obtained by other means then constraints on the rms radius, on the energy and on the number of bound states yield strong restrictions on the potential. Consequently, upper and lower limits on these quantities are interesting tools to obtain easily constraints on the interaction.

There exists a fairly large number of upper and lower limits on the energy of eigenstates in the literature \cite{duff47,schw61,calo67,barn76,lieb76,barn78,baum79,fern86,thir02} as well as on the number of bound states supported by central potentials \cite{barg52,schw61,calo65,glas76,mart77,lieb80,brau03a,brau03b}. Similar results concerning the rms radius are scarcer. A first general inequality gives a lower limit on the rms radius of a $\ell$-wave state in terms of the average kinetic energy ($\hbar=2m=1$) \cite[p. 73]{gros97}
\begin{equation}
\label{eq1}
\langle r^2\rangle \ge \frac{(2\ell+3)^2}{4\langle T\rangle}.
\end{equation}
This relation is however not very useful except in the case of power-law potentials, $V(r)=\text{sgn}(p) g r^p$ ($p>-2$), for which the virial theorem gives a simple relation between the energy, $E$, of the $\ell$-wave state and the average kinetic energy: $\langle T\rangle=pE/(p+2)$. In this case, the relation (\ref{eq1}) simply reads
\begin{equation}
\label{eq2}
\langle r^2\rangle \ge (2\ell+3)^2\, \frac{p+2}{4 p E}.
\end{equation}

In the case of a vanishing angular momentum, $\ell=0$, another restriction is given by the Bertlmann-Martin inequality \cite{bert80}
\begin{equation}
\label{eq3}
\langle r^2\rangle_{\ell=0} \le \frac{3}{E_{\ell=1}-E_{\ell=0}}.
\end{equation}
This simple relation is actually also applicable to systems composed by $N>2$ identical particles provided that no symmetry is required for the wave function \cite{bert81}. However, this formula yields restrictions only on the size of the ground state of the system and the energy of two levels needs to be known. 

In sec. \ref{sec2}, we propose several rigorous upper and lower limits on the rms radius applicable to systems composed by two-body bound by a central potential. Some of these limits are applicable to arbitrary eigenstates of the systems (any value of the radial quantum number, $n$, and of the angular momentum $\ell$). These limits involve the energy of the eigenstate considered as well as the potential itself. Upper and lower limits on the energy of the eigenstate can then be used to obtain upper and lower limits on the rms radius as a function of the potential only and to infer constraints on the values of its parameters. In sec. \ref{sec3}, as a simple application of the results obtained in this work, we use a lower bound on the rms radius to study weakly bound systems (quantum halo states) and to find a criteria for the occurrence of such states. Some tests of the other limits are reported in sec. \ref{sec4}. At last, we present some conclusions in sec. \ref{sec5}.

\section{Upper and lower limits on the rms radius}
\label{sec2}

To obtain various upper and lower limits on the rms radius, we consider the Schr\"odinger equation with a central potential
\begin{equation}
\label{eq4}
u''(r)=\left[V(r)+\frac{\ell(\ell+1)}{r^2}-E\right]u(r),
\end{equation}
where $u(r)=r R(r)$ and $R(r)$ is the radial wave function; the angular part is obviously given by the spherical harmonics $Y_{\ell m}(\theta,\varphi)$. For simplicity of the notations, we do not write the indices $n$ and $\ell$ on the function $u(r)$ and on the energy $E$, except if this is necessary for the clarity of the formula. The negative part of potential is supposed to be less singular than $r^{-2}$ at the origin and the potential is supposed to be piecewise continuous elsewhere. 

To obtain the relation from which each limit will be derived in this work, we multiply the relation (\ref{eq4}) by $r^2 u(r)$ and we integrate to obtain
\begin{equation}
\label{eq5}
\int_0^{\infty}dr\, r^2 u(r)\, u''(r)=\int_0^{\infty}dr\, r^2V(r)\, u(r)^2+\ell(\ell+1)-E\langle r^2\rangle.
\end{equation}
Integration by parts of the normalization condition of the wave function $u(r)$, $\int_0^{\infty}dr\, u(r)^2=1$, leads to
\begin{equation}
\label{eq6}
2\int_0^{\infty}dr\, r\, u(r)\, u'(r)=-1.
\end{equation}
In all cases throughout this paper, it is easy  to verify that the boundary terms of all integration by parts are vanishing if the energy $E$ and the rms radius $\langle r^2\rangle$ have a sense. Integration by parts of the left hand side of the relation (\ref{eq5}) together with (\ref{eq6}) leads to
\begin{equation}
\label{eq7}
-E\langle r^2\rangle+\ell(\ell+1)=1-\int_0^{\infty}dr\, r^2V(r)\, u(r)^2-\int_0^{\infty}dr\, r^2\, u'(r)^2.
\end{equation}

With the help of the equality (\ref{eq7}), it is possible to obtain various upper and lower limits on the rms radius. We gives these limits in the next sections.

\subsection{Simple upper limit}
\label{sec2.1}

The first upper limit we present is directly obtained from the relation (\ref{eq7}). Since $\int_0^{\infty}dr\, r^2 u'(r)^2 \ge 0$, we have
\begin{eqnarray}
\label{eq8}
-E\langle r^2\rangle&\le& 1-\int_0^{\infty}dr\, r^2V(r)\, u(r)^2-\ell(\ell+1)\nonumber \\
                    &\le& 1+ \sup_{0\le r<\infty}\left[-r^2V(r) \right]-\ell(\ell+1).
\end{eqnarray}
We still need to show that the right hand side of the last inequality of (\ref{eq8}) is finite, to have non trivial results, and positive for non vanishing value of the angular momentum. This last inequality is finite only if the negative part of the potential  decreases as $r^{-2}$ at infinity or faster. It is well known that the class of potentials characterized by a negative part which decrease faster than $r^{-2}$ at infinity support only a finite number of bound states \cite{cour53}. Consequently, there exist a maximal value, $L$, of the angular momentum, $\ell$, above which no bound state exists. If $L^+$ is defined by the relation
\begin{equation}
\label{eq9}
\sup_{0\le r<\infty}\left[-r^2V(r) \right]=\left(L^+ +\frac{1}{2}\right)^2,
\end{equation}
then it is also well known that $L^+$ is an upper limit on $L$, $L\le L^+$ \cite{cour53,brau03b}. The upper limit (\ref{eq8}) takes then the form
\begin{equation}
\label{eq10}
-E\langle r^2\rangle\le \frac{5}{4}+ (L^+)^2-\ell^2+L^+ -\ell,
\end{equation}
which proves that its right hand side is indeed positive. 

Potentials characterized by negative parts which decrease exactly as $cr^{-2}$ at infinity are at the borderline between potentials which possess a finite number of bound states and potentials for which this number is infinite. It has been proved that these potentials support only a finite number of bound states if $c$ is small enough \cite{stub90}. Thus in this case, there exists a maximal value, $L$, of the angular momentum. The upper limit $L^+$ is expected to be valid and the relation (\ref{eq10}) proves that the right hand side of (\ref{eq8}) is positive. However, to be completely rigorous, one should study this class of potentials in detail and prove that indeed that the upper limit $L^+$ is correct.

The simple upper limit (\ref{eq8}) shows that the product of the absolute value of the energy by the mean square radius cannot increase faster than linearly with the depth of the potential.

This upper limit will be significantly improved in the next section. We have presented this result because it is very simple and the discussion about the positivity of the right hand side of the inequality (\ref{eq8}) will be useful later.

\subsection{Main results}
\label{sec2.2}

The result (\ref{eq8}) obtained in the previous section can be improved provided we take into account the contribution of the last term of Eq. (\ref{eq7}). Such a contribution is easily obtained. An integration by parts leads to the equality
\begin{equation}
\label{eq11}
\int_0^{\infty}dr\, r^2 u'(r)^2=-\frac{2}{3}\int_0^{\infty}dr\, r^3 u'(r) u''(r).
\end{equation}
The second derivative of the wave function which appears in (\ref{eq11}) can be replaced using the Schr\"odinger equation (\ref{eq4}). A second integration by parts yield the desired result 
\begin{equation}
\label{eq12}
\int_0^{\infty}dr\, r^2 u'(r)^2=\frac{1}{3}\int_0^{\infty}dr\, (r^3 V(r))' \, u(r)^2+\frac{\ell(\ell+1)}{3}-E\langle r^2\rangle.
\end{equation}

The relation (\ref{eq7}) together with the identity (\ref{eq12}) leads to
\begin{equation}
\label{eq13}
-2E\langle r^2\rangle+\frac{4}{3}\ell(\ell+1)=1+\frac{1}{3}\int_0^{\infty}dr\, W(r)\, u(r)^2,
\end{equation}
where
\begin{equation}
\label{eq14}
W(r)=-(6 V(r)+r V'(r))r^2.
\end{equation}
The upper limit on the rms radius simply reads
\begin{equation}
\label{eq15}
-E\langle r^2\rangle\le \frac{1}{2}+\frac{1}{6}\sup\left[W(r)\right]-\frac{2}{3}\ell(\ell+1).
\end{equation}
This upper limit (\ref{eq15}) is non trivial if the negative part of the potential decreases as $r^{-2}$ at infinity or faster and if the positive part of the potential is less repulsive than $r^{-6}$ at the origin. The positivity of the right hand side of the upper limit (\ref{eq15}) is simple to prove. In sec. \ref{sec2.1}, we have obtained that the quantity $\sup\left[-r^2V(r) \right]-\ell(\ell+1)$ (see Eq. (\ref{eq8})) is positive. In contrast, we need now to show that the quantity $\sup\left[W(r)\right]/4-\ell(\ell+1)$ is positive. For this purpose it is sufficient to prove that the inequality $\sup\left[-r^2V(r) \right]\le \sup\left[W(r)\right]/4$ is verified. Suppose that the supremum of the function $-r^2V(r)$ is reached for $r=\bar{r}$. At this point, we have 
\begin{equation}
\label{eq16}
-\frac{1}{4}\bar{r}^3 V'(\bar{r})=\frac{1}{2}\bar{r}^2 V(\bar{r}).
\end{equation}
This infers that value of the supremum of $-r^2V(r)$ is equal to the value of $W(r)/4$ at this point: $-\bar{r}^2 V(\bar{r})=W(\bar{r})/4$. Thus the two functions have a crossing at $r=\bar{r}$ implying that the supremum of $W(r)/4$ cannot be smaller than the supremum of $-r^2V(r)$.

Similarly, the lower limit obtained from (\ref{eq13}) reads
\begin{equation}
\label{eq17}
-E\langle r^2\rangle\ge \frac{1}{2}+\frac{1}{6}\inf\left[W(r)\right]-\frac{2}{3}\ell(\ell+1).
\end{equation}
The inequality (\ref{eq17}) is non trivial if the positive part of the potential decrease as $r^{-2}$ at infinity or faster and is less singular than $r^{-2}$ or more singular than $r^{-6}$ at the origin. For potentials less singular than $r^{-2}$ at the origin, we have $W(0)=0$ yielding $\inf[W(r)]\le 0$. Consequently, the lower limit is only non trivial for S-wave states in this case. Moreover, if the infimum of $W(r)$ is negative, the lower limit becomes trivial for a depth of the potential large enough.

A case where the lower limit (\ref{eq17}) leads certainly to non trivial results concerns S-wave states and potentials such as $W(r)$ is \textit{non negative}. In this case we obtain the simple lower limit
\begin{equation}
\label{eq18}
-2E_{\ell=0}\langle r^2\rangle_{\ell=0}\ge 1.
\end{equation}
This relation is used in sec. \ref{sec3} to obtain a criteria for the occurrence of S-wave quantum halo states. We show that this criteria is still efficient even if the function $W(r)$ associated to the potential is somewhere slightly negative. Note also that the lower limit (\ref{eq18}) has been obtained previously but for a more restrictive class of potentials composed of potentials of finite range, $V(r>R)=0$, which are less singular than $r^{-2}$ at the origin \cite{lass98}.

It is interesting to note that, contrary to Eq.~(\ref{eq8}), the upper and lower limits (\ref{eq15}) and (\ref{eq17}) are best possible in the sense that there exists a potential which turn these inequalities into equalities. This potential is such as $W(r)=c$, where $c$ is a constant. The potential is then given by 
\begin{equation}
\label{eq18b}
V(r)=a\left(\frac{R}{r}\right)^6-b\left(\frac{R}{r}\right)^2,
\end{equation}
and $c=4 b R^2$.

\subsection{Upper limit for the lowest $\ell$-wave states}
\label{sec2.3}

To conclude this section devoted to the derivation of upper and lower limits on the rms radius, we present an upper limit applicable to the lowest $\ell$-wave states (no node in the wave function). This upper limit yields in general less restrictive constraints on the rms radius than the upper limits (\ref{eq8}) and (\ref{eq15}), except possibly for weakly bound systems.

We begin with the inequality
\begin{eqnarray}
\label{eq19}
r^2 u^2(r) &=&\left[\int_0^r dt\, (t u(t))'\right]^2\nonumber \\
                    &\le& r \left[\int_0^{\infty} dt\, [(t u(t))']^2\right]=r \int_0^{\infty}dt\, t^2 u'(t)^2,
\end{eqnarray}
where the Cauchy-Schwarz inequality is used. The inequality (\ref{eq19}) together with the relation (\ref{eq7}) yield
\begin{equation}
\label{eq20}
-E\langle r^2\rangle\le 1+{\cal I} \int_0^{\infty}dr\, r^2 u'(r)^2-\ell(\ell+1),
\end{equation}
where
\begin{equation}
\label{eq21}
{\cal I}=-1+\int_0^{\infty}dr\, r V^-(r),
\end{equation}
with $V^-(r)=\max(0,-V(r))$. The Jost-Pais necessary condition \cite{jost51} implies that ${\cal I}$ is positive if the potential supports at least one bound state. The quantity ${\cal I}$ does not diverge if the potential decrease faster than $r^{-2}$ at the infinity. Now, we consider the following inequality 
\begin{eqnarray}
\label{eq22}
u'(r) &=&-\int_r^{\infty} dt\, u''(t), \nonumber \\ 
&=&-\int_r^{\infty} dt\, \left[V(t)+\frac{\ell(\ell+1)}{t^2}-E\right]u(t), \nonumber \\
                    &\le& \int_r^{\infty}dt\, V^-(t) u(t) +\int_r^{\infty} dt\, \left[E-\frac{\ell(\ell+1)}{t^2}\right]u(t), \nonumber \\
&<& \int_r^{\infty}dt\, V^-(t) u(t).
\end{eqnarray}
The last inequality in Eq. (\ref{eq22}) is valid only for wave functions without node, i.e. the lowest $\ell$-wave states. The Cauchy-Schwarz inequality together with the inequality (\ref{eq22}) yield
\begin{eqnarray}
\label{eq23}
u'^2(r) &<&\int_r^{\infty} dt\, [V^-(t)]^2 \int_r^{\infty} dt\, u(t)^2, \nonumber \\
                    &<& \int_r^{\infty} dt\, [V^-(t)]^2.
\end{eqnarray}
We obviously suppose that the integral, from 0 to $\infty$, of the square of the negative part of the potential exists. This last inequality can be used with the relation (\ref{eq20}), and after an integration by parts we obtain
\begin{equation}
\label{eq24}
-E\langle r^2\rangle< 1+\frac{{\cal I}}{3} \int_0^{\infty}dr\, r^3 [V^-(r)]^2-\ell(\ell+1),
\end{equation}
where ${\cal I}$ is defined by (\ref{eq21}). The right hand side of the inequality (\ref{eq24}) behaves as third power of the strength of the potential. Clearly, for large values of the strength this upper limit yields poor constraint on the rms radius but for values of the strength close to the critical value (value at which a first $\ell$-wave bound state appears), and if this critical value is small enough, we show in sec. \ref{sec4} that this upper limit is better than those reported in secs. \ref{sec2.1} and \ref{sec2.2}.

\section{Criteria for occurrence of halos}
\label{sec3}

In this section, we use the lower limit (\ref{eq18}) to obtain a criteria for the occurrence of S-wave quantum halo states. These states are threshold phenomena characterized by large mean square radii and small binding energies. They occur in nuclear physics as halo nuclei (see for example \cite{bert93,hans93,riis94,hans95,jens04}) and in molecular physics as weakly bound dimers (see for instance \cite{luo93,scho94,luo96}). For quantum halo states, the separation energy of the two bodies of the system (a nucleus and a nucleon or two atoms) is much smaller than the mean binding energy of the particles which compose eventually these bodies. These many-body systems can then be treated as two-body systems interacting through a potential. 

The idea of the criteria is the following: knowing the two bodies which compose the system as well as the central interaction, we determine for which binding energies quantum halo states exist. The application presented in this section is somewhat complementary to previous studies found in the literature (see for example \cite{riis92,fedo93,jens00,riis00,lomb02}).

Quantum halo states are characterized by an extension far out into the classical forbidden region. We consider that a state with a binding energy $E$ is a halo state if its rms radius is larger than its classical radius 
\begin{equation}
\label{eq25}
\langle r^2\rangle^{1/2} \ge \sigma \, r_0,
\end{equation}
with $E=V(r_0)$. The value of $\sigma$ can be estimated if we consider that in a quantum halo states the probability to find the particles with a interdistance greater than the classical interdistance is greater than $50 \%$ (see for example \cite{riis94,hans95}). The value of $\sigma$ obtained with this definition varies from $1.37$ for the square well to $1.68$ for a potential which decrease as $r^{-3}$ at the infinity. In this work we take $\sigma=2$ in all numerical calculations for simplicity. 

With the lower limit (\ref{eq18}) and the constraint (\ref{eq25}) we can write 
\begin{equation}
\label{eq26}
\langle r^2\rangle \ge \frac{1}{-2E}= \frac{1}{-2V(r_0)}\ge \sigma^2 \, r_0^2.
\end{equation}
Thus we are in the halo regime if $-2\sigma^2 \, r_0^2 V(r_0)\le 1$.

We can now obtain the criteria for the occurrence of halo states. We write the potential in the convenient form $V(r)=-gR^{-2} v(r/R)$. Notice that the number of bound states in the potential and the critical value of the strength $g$ at which new bound states appear do not depend on $R$. We first search for the largest solution, $x_0$, of the equation
\begin{subequations}
\label{eq27}
\begin{equation}
\label{eq27a}
x_0^2\, v(x_0)=\frac{1}{2\sigma^2\, g^N_{\text{c}}},
\end{equation}
where $x_0=r_0/R$ and where the coupling constant $g$ is replaced by its critical value $g^N_{\text{c}}$ for which the $N$th eigenstate has a vanishing binding energy. Since $g \ge g^N_{\text{c}}$, we obtain a value of $x_0$ slightly underestimated which lead to a slightly lower value of the energy $E_H$ above which quantum halo states appears. This effect will be attenuated (suppressed in practice) if we use upper limits on $g^N_{\text{c}}$. Accurate upper and lower limits on the value of $g^1_{\text{c}}$ and on $g^{N>1}_{\text{c}}$, involving only the potential, can be found in the literature \cite{barg52,schw61,calo65,glas76,mart77,lieb80,brau03a,brau03b,brau03c,brau04a,brau04b}. 

The energy $E_H$ is then given by
\begin{equation}
\label{eq27b}
E_H=-g^N_{\text{c}}R^{-2}v(x_0)=-\frac{R^{-2}}{2(\sigma \, x_0)^2} .
\end{equation}
\end{subequations}
Consequently, states characterized by a binding energy $E$ larger than $E_H$ are characterized by a rms radius satisfying the inequality (\ref{eq25}). 

Relevant information about occurrence of quantum halo sates is obtained with the criteria (\ref{eq27}) only if the energies at which these states appear is obtained with a reasonable accuracy. In other words, the inequality (\ref{eq25}) should be verified and reasonably close to saturation. Two tests are performed below.
 
Several remarks are in order. 
\begin{itemize}
\item
Since the inequalities $g^1_{\text{c}}<g^2_{\text{c}}<\ldots<g^{N}_{\text{c}}$ are always verified, $x_0$ is larger for an excited state than for the ground state (see (\ref{eq27a})). Consequently, the energy $E_H$ is greater for excited states than for ground states. This clearly indicates that halo states are likely to be ground states instead of excited states. This conclusion may be incorrect for potentials which vanish identically beyond a given radius $x^*=r^*/R$. In this case, $x_0$ could stay constant for all values of $N$ if the radius $x^*$ is small enough. This is the case for a square-well potential as discussed below. 
\item From the relation (\ref{eq27b}), halos states have best chance to appear in potentials with a small range $R$. Indeed, when $R$ varies, the quantities $g^N_{\text{c}}v(x_0)$ or $x_0$ remain unchanged. This result is simple to understand: both $\langle r^2\rangle^{1/2}$ and $r_0$ scale like $R$, consequently, their ratio is independent of $R$ but the energy $E$ scales like $R^{-2}$. Consequently, the energy for which the system is characterized by a given value of the ratio $\langle r^2\rangle^{1/2}/r_0$ scales like $R^{-2}$.
\item When the two constituents of the halo are characterized by a finite size, like atoms or nuclei, halo states have best chance to appear for small sizes and for small reduced masses of these constituents (if a repulsion exist for small interdistance).
\end{itemize}

To illustrate the last affirmation, we consider the following interaction
\begin{equation}
\label{eq28}
V(r)=gR^{-2}\left[\left(\frac{R}{r}\right)^{2(n-1)}-\left(\frac{R}{r}\right)^n\right].
\end{equation}
The repulsive part of the potential takes roughly into account the Pauli repulsion and $R$ is then linked with the sizes of the particles interacting through this potential. The attractive part describe various kind of interactions depending on the value of $n$. The interaction of a charge and an induced dipole correspond to $n=4$; both $n=6$ and $n=7$ correspond to van der Waals forces, of London and Casimir-Polder type, respectively. For $n=6$, we have a Lennard-jones $(10,6)$ potential. The particular form of the interaction (\ref{eq28}) is simply chosen to allow analytical calculations and is not intended to describe physical systems accurately but instead to get insights for the gross characteristics of some physical systems. 

The function $W(r)$ (see (\ref{eq14})) is positive for $n\ge 4$ and the lower limit (\ref{eq18}) and then the criteria (\ref{eq27}) can be used. An estimation of the critical value $g^{N}_{\text{c}}$ of the strength $g$ is obtained with the formula 
\begin{equation}
\label{eq29}
g^{N}_{\text{c}}\le (N\pi)^2 \left[\int_0^{\infty}dx\, \sqrt{v^-(x)}\right]^{-2},
\end{equation}
which is applicable for $n\ge 4$ \cite{brau04a}. This leads to 
\begin{equation}
\label{eq30}
g^{N}_{\text{c}}\le (2N (n-2))^2.
\end{equation}
The quantity $x_0$ is obtained easily
\begin{equation}
\label{eq31}
x_0^{n-2}=\frac{1+\sqrt{1-4\gamma}}{2\gamma}\cong \frac{1}{\gamma}-1,
\end{equation}
where $\gamma=1/(2\sigma^2 g^{N}_{\text{c}})\le 0.0078$ which justify the expansion around $\gamma=0$. The square root in the relation (\ref{eq31}) imposes $g^{N}_{\text{c}} \ge 1/2$. This is always verified since the simple lower bound $g^{N}_{\text{c}}\ge 2 N (n-2)$ can be obtained with the Bargmann-Schwinger inequality \cite{barg52,schw61}. The relations (\ref{eq27b}), (\ref{eq30}) and (\ref{eq31}) lead to
\begin{equation}
\label{eq32}
E_H=-\frac{\left(2^{n+4}\sigma^{2n}N^2(n-2)^4\right)^\frac{-1}{n-2}}{2\mu R^2},
\end{equation}
where the reduced mass of the system, $\mu$, has been written explicitely. The energy $E_H$ decreases when $n$ grows giving more chance to halo states to exist. For fixed value of $n$, $E_H$ increases with $\mu$ and $R$ and also with $N$. 

We can calculate $E_H$ with the formula (\ref{eq32}) for $n=6$ and $R=2.640$~\AA \cite{hirs54}. This model could be used to describe roughly the helium dimer $^4$He$_2$. We find that $E_H=-0.82$~$\mu$eV, while the experimental binding energy is found to be around $-0.095$~$\mu$eV \cite{gris00}. In this model, the helium dimer is a halo state. Actually the value of $R$ used in this simple calculation was adjusted for a Lennard-Jones $(12,6)$ potential \cite{hirs54}. If this last potential is used, instead of a Lennard-Jones $(10,6)$ potential, we find a slightly modified value $E_H=-0.89$~$\mu$eV.

We can also study how the energy $E_H$ is sensitive to the asymptotic behavior of the potential. We simply choose the potential
\begin{equation}
\label{eq32b}
V(r)=-\frac{gR^{-2}}{1+(r/R)^n}.
\end{equation}
For $n\ge 4$, the formula (\ref{eq29}) can be used to obtain an upper bound on the critical coupling constant. We have
\begin{equation}
\label{eq32c}
g^{N}_{\text{c}}\le \frac{N^2 \pi^3}{\Gamma^2\left(\frac{1}{2}-\frac{1}{n}\right)\Gamma^2\left(1+\frac{1}{n}\right)},
\end{equation}
where $\Gamma(x)$ is the Euler gamma function. To find the quantity $x_0$, we need to solve
\begin{equation}
\label{eq32d}
x_0^n=\frac{x_0^2}{\gamma}-1 \cong \frac{x_0^2}{\gamma},
\end{equation}
with $\gamma = 1/(2\sigma^2 g^{N}_{\text{c}})$. This leads to
\begin{equation}
\label{eq32e}
x_0\cong (2\sigma^2 g^{N}_{\text{c}})^{\frac{1}{n-2}}.
\end{equation}
The energy at which halo states appear is given by
\begin{eqnarray}
\label{eq32f}
E_H&\cong& -R^{-2}(2\sigma^2)^{\frac{n}{2-n}}(g^{N}_{\text{c}})^{\frac{2}{2-n}}, \\
  &\cong& -R^{-2} s(N,n). 
\end{eqnarray}

Before to draw some conclusions about the influence of the asymptotic behavior of the potential on the energy $E_H$, we mention that, as expected, the rms radius satisfy the inequality (\ref{eq25}) if the binding energy is larger than $E_H$. If the binding energy is chosen to be equal to $E_H$ for $N=1$, an exact numerical calculation shows that the value of the ratio of the rms radius over the classical radius is, for example, $2.44$ for $n=5$, $2.47$ for $n=10$, $2.44$ for $n=20$, $2.41$ for $n=50$ and $2.40$ for $n=100$. The same calculation for $N=2$ leads the following values for the same ratio: $2.29$ for $n=5$, $2.38$ for $n=10$, $2.42$ for $n=20$, $2.40$ for $n=50$ and $2.40$ for $n=100$. These results indicate that even if the function $W(r)$ computed with the potential (\ref{eq32b}) is partially negative for $n>6$, the criteria stays applicable for large values of $n$ since the inequality (\ref{eq25}) is always verified.

\begin{figure}
\includegraphics*[width=10cm]{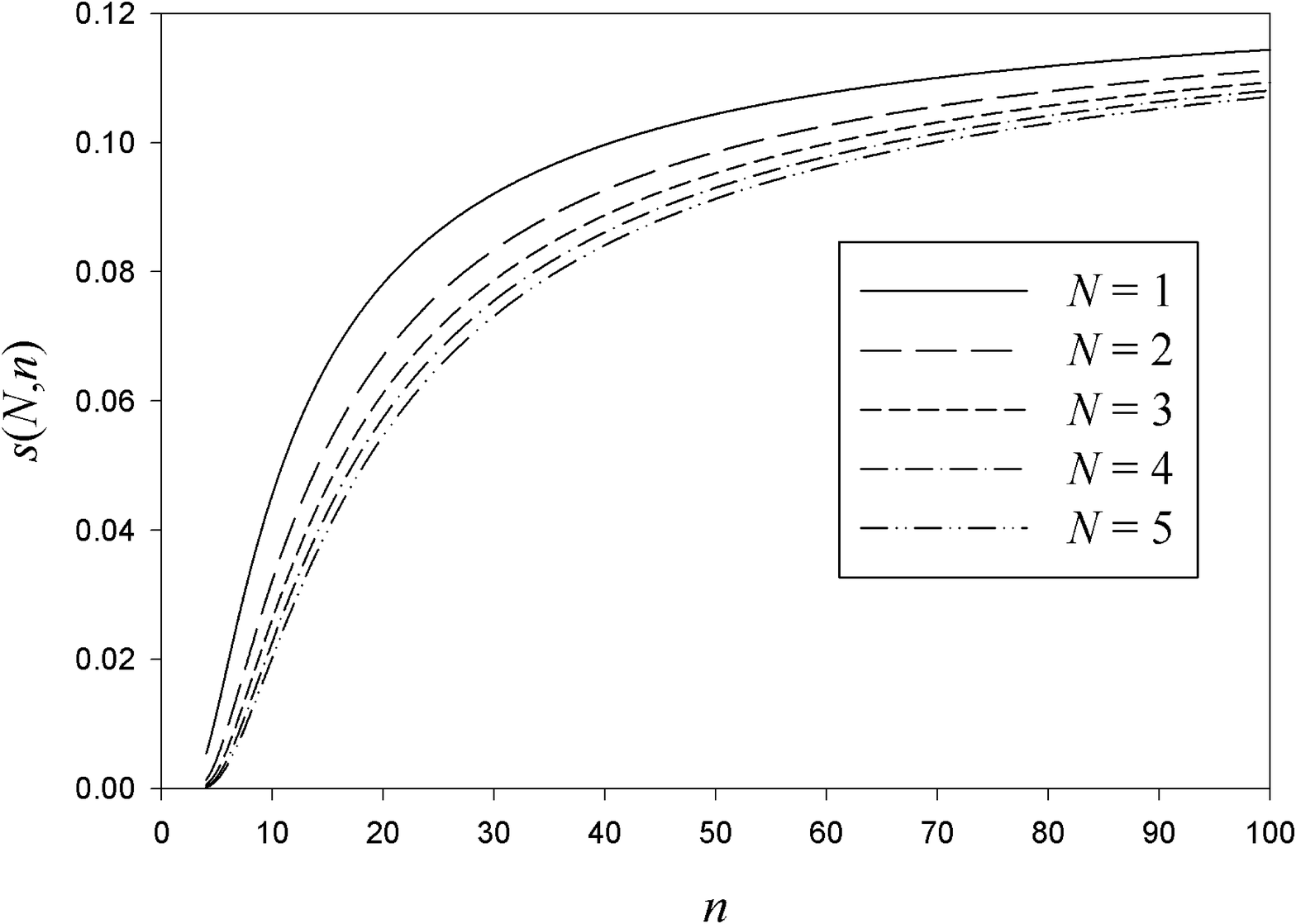}
\caption{\label{fig1}$s(N,n)$ as a function of $n$ for several values of $N$.}
\end{figure}

Fig. \ref{fig1} indicates clearly that halos could exist more easily when $n$ is large since in this case the energy at which they appear, $E_H$, is smaller (keeping $R$ constant obviously). It is also clear that halos have more chance to exist as ground states than as excited states. However this last conclusion is only partially correct since we have  $\lim_{n\rightarrow \infty} s(N,n) = 1/(2\sigma^2)$ for all value of $N$. In this limit the potential (\ref{eq32b}) reduces to a square-well potential. This property can be verified with an exact calculation, i.e. for a square-well potential, the energy at which halo states appears is almost the same for the ground state and for excited states. For example, with the potential  $V(r)=-V_0 \exp(-r/R) \theta(R-r)$, halo states have even marginally more chance to exist as excited state than as ground states. It is also clear on the Fig.~\ref{fig1} that the ratio of the energy $E_H$ for the ground states over the energy $E_H$ for the first excited states, $E_H(N=1)/E_H(N=2)$, decreases as $n$ grows. The values of this ratio are for example: $4$ for $n=4$, $2$ for $n=6$ and $1.41$ for $n=10$.

The same study could be performed with the potential $V(r)=-g R^{-2}\exp(-(r/R)^n)$. The same qualitative behaviors of $E_H$ than those obtained with the potential (\ref{eq32b}) are observed. The ratio $E_H(N=1)/E_H(N=2)$ also decreases as $n$ grows but the values of this ratio are now smaller: $2.15$ for $n=1$, $1.35$ for $n=2$ and $1.12$ for $n=5$.

We can also study how the energy $E_H$ is sensitive to the details of the repulsive potential near the origin. We simply choose the potential
\begin{equation}
\label{eq32g}
V(r)=gR^{-2}\left(\left(\frac{R}{r}\right)^n-\left(\frac{R}{r}\right)^4\right).
\end{equation}
For this potential, the function $W(r)$ is positive for $n\ge 6$.
For $n\ge 5$, the formula (\ref{eq29}) can be used to obtain an upper bound on the critical coupling constant. We obtain
\begin{equation}
\label{eq32h}
g^{N}_{\text{c}}\le 4 N^2 \pi (n-4)^2\Gamma^2\left(\frac{3n-10}{2n-8}\right)\Gamma^{-2}\left(\frac{1}{n-4}\right),
\end{equation}
To find the quantity $x_0$, we need to solve
\begin{equation}
\label{eq32i}
\gamma x_0^2=1-x_0^{4-n}\cong 1,
\end{equation}
with $\gamma = 1/(2\sigma^2 g^{N}_{\text{c}})$. This leads to
\begin{equation}
\label{eq32j}
x_0\cong (2\sigma^2 g^{N}_{\text{c}})^{1/2}.
\end{equation}
The energy at which halo states appear is given by
\begin{eqnarray}
\label{eq32k}
E_H&\cong& -R^{-2}\frac{1}{(2\sigma^2)^2 g^{N}_{\text{c}}}, \\
  &\cong& -R^{-2} t(N,n). 
\end{eqnarray}

\begin{figure}
\includegraphics*[width=10cm]{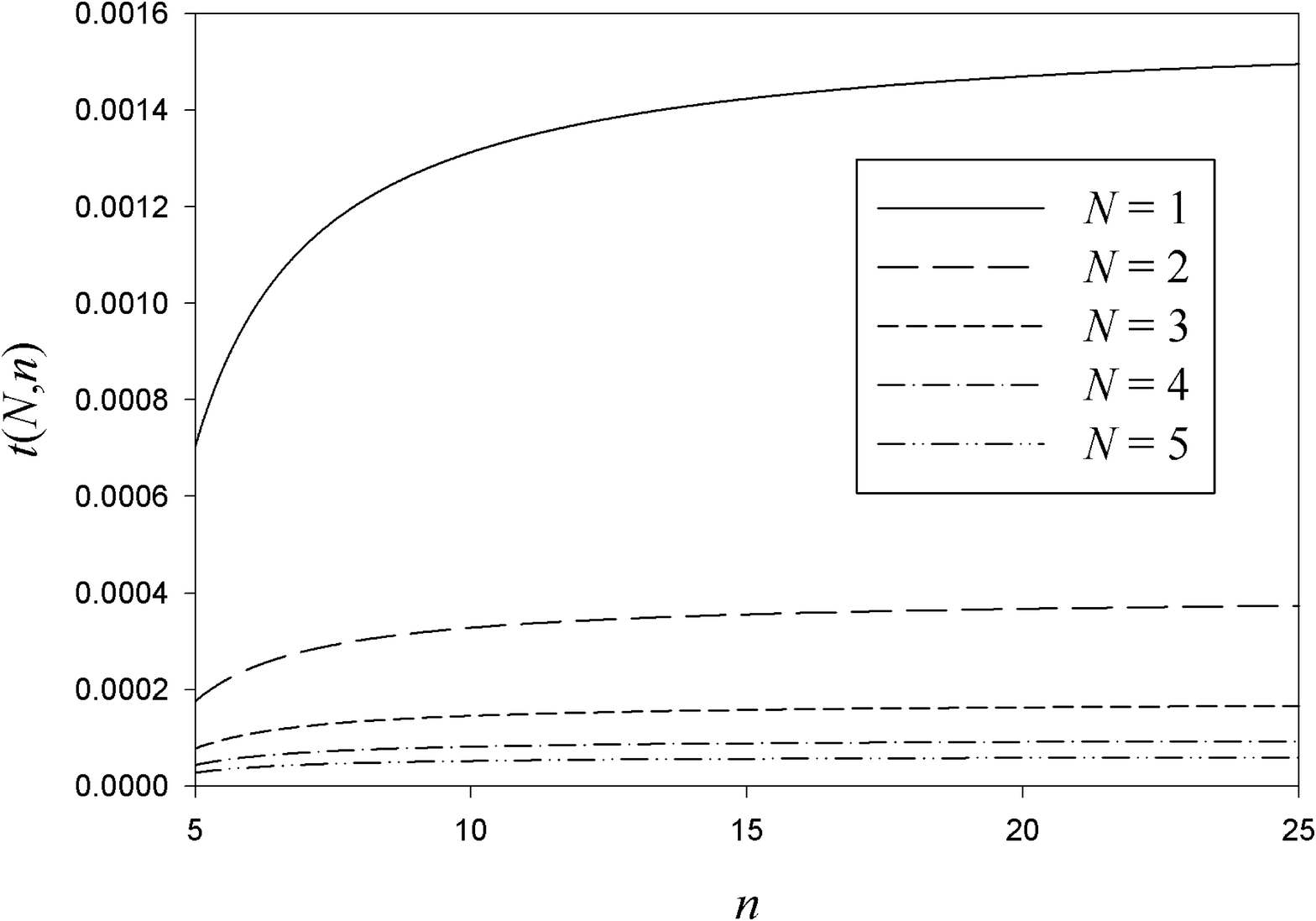}
\caption{\label{fig2}$t(N,n)$ as a function of $n$ for several values of $N$.}
\end{figure}

Fig. \ref{fig2} indicates that existence of halo states is not very sensitive to the details of the repulsive potential. For the ground state, the energy $E_H$ increases by a factor $2$ for $n$ going from $5$ to $25$ whereas in the previous case, see Fig. \ref{fig1}, the increase was by a factor $10$ in the same interval of $n$.

Application to nuclear halos is possible. For this purpose we consider a Wood-Saxon potential as interaction between the nucleon and the nucleus
\begin{equation}
\label{eq33}
V(r)=-\frac{V_0}{1+\exp((r-R)/a)},
\end{equation}
with $V_0=g a^{-2}$, $R=r_0 A^{1/3}$, $r_0=1.27$ fm and $a=0.67$ fm \cite{bohr69}. For this potential, the function $W(r)$ is (slightly) partially negative whereas the criteria rigorously apply only to potential yielding a function $W(r)$ everywhere positive. In Table \ref{tab1}, we report a comparison between exact results and results obtained with the formulas (\ref{eq27}) which prove that the criteria is applicable. 

\begin{table}[htb]
\protect\caption{Comparison between the energy $E_H$ given by (\protect\ref{eq36}) with the exact energy $E_{\text{ex}}$ at which at which $\langle r^2\rangle^{1/2} = 2 \, r_0$ for several values of the atomic number $A$. The ratio $\langle r^2\rangle^{1/2}/r_0$ at energy $E_H$ is also given. The energies are given in MeV.}
\label{tab1}
\begin{center}
\begin{tabular}{cccc}
\hline\noalign{\smallskip}
$A$ & $-E_{\text{ex}}$ & $-E_H$ & $\frac{\langle r^2\rangle^{1/2}}{r_0}$   \\
\noalign{\smallskip}\hline\noalign{\smallskip}
1   &  0.30  &  0.26  &  2.11     \\
5   &  0.14  &  0.11  &  2.20     \\
10  &  0.11  &  0.087 &  2.23     \\
15  &  0.10  &  0.075 &  2.25     \\
20  &  0.097 &  0.067 &  2.27     \\
25  &  0.089 &  0.062 &  2.28     \\
50  &  0.069 &  0.047 &  2.30     \\
100 &  0.053 &  0.035 &  2.32     \\
\noalign{\smallskip}\hline
\end{tabular}
\end{center}
\end{table}

The critical coupling constant of the potential (\ref{eq33}) is simply estimated with the WKB formula
\begin{equation}
\label{eq34}
g^{N}_{\text{c}}\cong \left[\frac{(N-1/4)\pi}{\alpha+2\ \text{arcsinh} 1}\right]^2,
\end{equation}
with $\alpha=R/a\cong 1.9\, A^{1/3}$. The quantity $x_0$ is obtained with the Lagrange inversion formula
\begin{equation}
\label{eq35}
x_0\cong \alpha\,\left(\alpha^2/ \gamma-1 \right)^{1/\alpha},
\end{equation}
with $\gamma = 1/(2\sigma^2 g^{1}_{\text{c}})$. We restrict the calculation to the case $N=1$ for which the formula (\ref{eq35}) is quite accurate. Taking into account the reduced mass of the system, we finally obtain
\begin{equation}
\label{eq36}
E_H\cong -1.6\ h(A)\, A^{-2/3}\, \text{MeV},
\end{equation}
where
\begin{equation}
\label{eq37}
h(A)=\frac{A+1}{A\,\left(\alpha^2/ \gamma-1 \right)^{2/\alpha}}.
\end{equation}
For a square-well potential, the energy $E_H$ scales almost exactly as $A^{-2/3}$, for a Wood-Saxon potential there is a correction. Indeed, taking the limit $n\rightarrow \infty$ of the relation (\ref{eq32f}), and taking into account the reduced mass of the system, we obtain for a square-well $E_H\cong -1.6\, [(A+1)/A]\, A^{-2/3}$ MeV (in agreement with a previous result \cite{fedo93}). The function $h(A)$ is slightly increasing ($h(1)\cong 0.16$ and $h(225)\cong 0.55$) and the modulus of the energy $E_H$ decreases slower than $A^{-2/3}$. A numerical fit of the formula (\ref{eq36}) leads to $E_H\cong -0.22\ A^{-2/5}$ MeV. This last expression depends obviously on the value of $\sigma$. If instead of $\sigma=2$ we choose $\sigma=1.5$, the fit becomes $E_H\cong -0.25\ A^{-1/2}$ MeV.

In Table~\ref{tab1}, to test the criteria (\ref{eq27}) applied to nuclear halos, we compare the value of the energy $E_H$ given by the formula (\ref{eq36}) with the exact energy $E_{\text{ex}}$ at which $\langle r^2\rangle^{1/2} = \sigma \, r_0$, with $\sigma=2$. We also give the exact value of the ratio $\langle r^2\rangle^{1/2}/r_0$ at energy $E_H$. This ratio is almost constant, it increases slowly with the atomic number $A$. In any case, the inequality (\ref{eq25}) is satisfied.

\section{Tests}
\label{sec4}

In this section we propose to test of the various upper and lower limits reported in sec.~\ref{sec2} with two simple potentials
\begin{equation}
\label{eq38}
V_1(r)=-\frac{g R^{-2}}{1+(r/R)^3},
\end{equation}
and
\begin{equation}
\label{eq39}
V_2(r)=-g R^{-1}\frac{\exp(-r/R)}{r}.
\end{equation}

\begin{table}[htb]
\protect\caption{Comparison between the exact value of the quantity $-E\langle r^2\rangle$ for the ground state and various upper and lower limits reported in sec.~\protect\ref{sec2} for the potential $V_1(r)$ and for $\ell=0$.}
\label{tab2}
\begin{center}
\begin{tabular}{cccccc}
\hline\noalign{\smallskip}
$g$ & $-E\langle r^2\rangle$ & Eq.~(\protect\ref{eq8}) & Eq.~(\protect\ref{eq15}) & Eq.~(\protect\ref{eq17}) & Eq.~(\protect\ref{eq24}) \\
\noalign{\smallskip}\hline\noalign{\smallskip}
1.35  &  0.51418  &  1.7143  &  1.0063 & 0.5 & 1.1549   \\
1.4   &  0.55325  &  1.7407  &  1.025  & 0.5 & 1.1825   \\
1.5   &  0.61348  &  1.7937  &  1.0625 & 0.5 & 1.2460   \\
1.75  &  0.74037  &  1.9259  &  1.1563 & 0.5 & 1.4592   \\
2     &  0.85531  &  2.0582  &  1.25   & 0.5 & 1.7623   \\
3     &  1.26746  &  2.5873  &  1.625  & 0.5 & 4.1773   \\
4     &  1.63531  &  3.1164  &  2      & 0.5 & 9.2480   \\
5     &  1.97341  &  3.6455  &  2.375  & 0.5 & 17.949   \\
\noalign{\smallskip}\hline
\end{tabular}
\end{center}
\end{table}

\begin{table}[htb]
\protect\caption{Same as Table \protect\ref{tab2} but for $\ell=1$.}
\label{tab3}
\begin{center}
\begin{tabular}{ccccc}
\hline\noalign{\smallskip}
$g$ & $-E\langle r^2\rangle$ & Eq.~(\protect\ref{eq8}) & Eq.~(\protect\ref{eq15}) & Eq.~(\protect\ref{eq24}) \\
\noalign{\smallskip}\hline\noalign{\smallskip}
6.945  &  0.06045  &  2.6746  &  1.771   & 46.942   \\
6.95   &  0.09321  &  2.6773  &  1.7729  & 47.045   \\
7      &  0.23641  &  2.7037  &  1.7917  & 48.142   \\
7.5    &  0.74323  &  2.9683  &  1.9792  & 59.982   \\
8      &  1.0652   &  3.2328  &  2.1667  & 73.583   \\
9      &  1.5923   &  3.7619  &  2.5417  & 106.55   \\
10     &  2.0529   &  4.291   &  2.9167  & 148.03   \\
\noalign{\smallskip}\hline
\end{tabular}
\end{center}
\end{table}

In Tables \ref{tab2} and \ref{tab3}, we compare the exact value of the quantity $-E\langle r^2\rangle$, for the potential $V_1(r)$ and for $\ell=0$ and $1$ respectively, with the various upper and lower limits reported in sec. \ref{sec2}. As expected the upper limit (\ref{eq15}) yields the strongest restrictions and is always better than the formula (\ref{eq8}) by construction. However for $\ell=1$, the constraints are poor for states characterized by a weak binding energy. As explain in sec. \ref{sec2.3}, the upper limit (\ref{eq24}) yields its best restrictions for the coupling constant $g$ close to the critical value $g_{\text{c}}$ ($1.3326 < g_{\text{c}} < 1.3403$ \cite{glas76,brau04b}). Useful restrictions are obtained only if this critical value is small enough. For $\ell=1$, we have $g_{\text{c}}\cong 6.94$ ($6.9221 < g_{\text{c}} < 6.9535$ \cite{glas76,brau04b}) and the relation (\ref{eq24}) gives poor restrictions. If $g_{\text{c}}$ is smaller than 1, the upper limit (\ref{eq24}) could be better than (\ref{eq15}) for $g$ close enough to $g_{\text{c}}$; this is the case for the potential $V(r)=-g R^{-3} r \exp(-r/R)$. For the potential $V_1(r)$, the lower limit (\ref{eq17}) yields non trivial results only for $\ell=0$. In this case, since the infimum of $W(r)$ is equal to $0$, the lower limit is a constant and is quite restrictive for values of $g$ close to the critical value $g_{\text{c}}$. In general, this lower limit yields strong restrictions when the binding energy of the system is small. This explain the good accuracy of the criteria obtained in sec.~\ref{sec3}.  

In Tables \ref{tab4} and \ref{tab5}, we compare the exact value of the quantity $-E\langle r^2\rangle$, for the potential $V_2(r)$ and for $\ell=0$ and $1$ respectively, with the various upper and lower limits reported in sec. \ref{sec2}. For this potential, the upper limit (\ref{eq24}) is not applicable and the lower limit (\ref{eq17}) yields non trivial results only for $\ell=0$. Again, as expected the upper limit (\ref{eq15}) yields the strongest restrictions and is obviously always better than the formula (\ref{eq8}). The lower limit (\ref{eq17}) is again very accurate for weak binding energy.

\begin{table}[htb]
\protect\caption{Same as Table \protect\ref{tab2} but for the potential $V_2(r)$ and $\ell=0$.}
\label{tab4}
\begin{center}
\begin{tabular}{ccccc}
\hline\noalign{\smallskip}
$g$ & $-E\langle r^2\rangle$ & Eq.~(\protect\ref{eq8}) & Eq.~(\protect\ref{eq15}) & Eq.~(\protect\ref{eq17}) \\
\noalign{\smallskip}\hline\noalign{\smallskip}
1.7   &  0.50948  &  1.6254  &  0.92778 & 0.49575    \\
1.75  &  0.53314  &  1.6438  &  0.94036 & 0.49563    \\
1.8   &  0.55699  &  1.6622  &  0.95294 & 0.4955     \\
1.9   &  0.60473  &  1.699   &  0.97810 & 0.49525    \\
2     &  0.65199  &  1.7358  &  1.0033  & 0.495      \\
3     &  1.0558   &  2.1036  &  1.2549  & 0.4925     \\
4     &  1.3429   &  2.4715  &  1.5065  & 0.49       \\
5     &  1.55389  &  2.8394  &  1.7582  & 0.4875     \\
\noalign{\smallskip}\hline
\end{tabular}
\end{center}
\end{table}

\begin{table}[htb]
\protect\caption{Same as Table \protect\ref{tab2} but for the potential $V_2(r)$ and $\ell=1$.}
\label{tab5}
\begin{center}
\begin{tabular}{cccc}
\hline\noalign{\smallskip}
$g$ & $-E\langle r^2\rangle$ & Eq.~(\protect\ref{eq8}) & Eq.~(\protect\ref{eq15})  \\
\noalign{\smallskip}\hline\noalign{\smallskip}
9.085  &  0.03585   &  2.3422  &  1.4528   \\
9.1    &  0.092294  &  2.3477  &  1.4565    \\
9.5    &  0.53225   &  2.4949  &  1.5572   \\
10     &  0.85248   &  2.6788  &  1.683   \\
11     &  1.3309    &  3.0467  &  1.9346     \\
12     &  1.7085    &  3.4146  &  2.1863    \\
13     &  2.0267    &  3.7824  &  2.4379        \\
14     &  2.3031    &  4.1503  &  2.6895    \\
15     &  2.5476    &  4.5182  &  2.9412    \\
\noalign{\smallskip}\hline
\end{tabular}
\end{center}
\end{table}

\section{Conclusions}
\label{sec5}

In this article, several rigorous upper and lower limits on the rms radius have been obtained for systems governed by central potentials. Some of these limits are applicable to eigenstates with arbitrary radial quantum number and angular momentum. Some of these limits yield in general strong restrictions on the rms radius as shown in sec.~\ref{sec4}. The simple lower limit (\ref{eq18}) is used to obtain a criteria for the occurrence of S-wave halo states. This criteria gives the binding energy, called $E_H$, above which the eigenstate is characterized by a large rms radius compared to the classical radius, $\langle r^2\rangle^{1/2} \ge \sigma \, r_0$, and is thus qualified as quantum halo states. The relevance of the criteria has been tested with various potentials and we found that accurate information are obtained. It is worth noting that the various formula derived in sec.~\ref{sec3} are applicable for arbitrary values of $\sigma$ but the various numerical values reported in the text and in the tables are computed for $\sigma=2$ for simplicity. This value could be changed for practical uses. However, the conclusions obtained in this work, and summarized below, do not depend on the precise value of $\sigma$ contrary, of course, to the energy $E_H$.

With this criteria we have shown that halo states are likely to be ground states and not radial excitations (except possibly for potentials which vanish identically beyond a given radius, like the square-well potential, for which the converse could be true). This conclusion complete others results obtained previously which proved that halo states are S- or P-wave states. We have shown that, when the two constituents of the halo are characterized by a finite size, like atoms or nuclei, halo states have best chance to appear for small sizes and for small reduced masses of these constituents (if a repulsion exist for small interdistance). The criteria is also used to confirm that halo state have more chance to exist in potentials which tends rapidly to zero asymptotically. We have also shown that, if the potential has a repulsive part near the origin, the existence of halo is not very sensitive to the detail of this repulsive part.

\begin{acknowledgments}
This work was supported by the National Funds for Scientific Research (FNRS), Belgium.
\end{acknowledgments}

\end{document}